\documentclass{article}

\def\xxinput#1{\input#1}

\xxinput{vsolj02.sty}

\usepackage{longtable}
\usepackage[dvipdfmx]{graphicx}
\usepackage[comma,colon]{natbib}

\usepackage[OT2,T1]{fontenc}

\def\cite{\citealt}
\setcitestyle{aysep={}}

\def\Nakataprep{C. Nakata et al. in preparation}
\def\Tampoprep{Y. Tampo et al. in preparation}

\def\commenta{$^*$}
\def\commentb{$^\dagger$}
\def\commentc{$^\ddagger$}
\def\commentd{$^\S$}
\def\commente{$^\|$}

\def\mvsh{$M_V(\rm SH)$}
\def\aesh{$A_{\rm ESH}$}

\begin{document}

\title{Emerging ordinary superhumps as the standard candle for WZ Sge stars}

\author{Taichi Kato$^1$}
\author{$^1$ Department of Astronomy, Kyoto University,
       Sakyo-ku, Kyoto 606-8502, Japan}
\email{tkato@kusastro.kyoto-u.ac.jp}

\begin{abstract}
\xxinput{abst.inc}
\end{abstract}

\section{Introduction}

   WZ Sge stars are a subclass of dwarf novae and they
usually show rare (typically once in a decade) and
large-amplitude (typically 8~mag) superoutbursts
[for general information of cataclysmic variables and
dwarf novae, see e.g. \citet{war95book}].
Although WZ Sge stars were originally defined as
dwarf novae showing rare, large-amplitude superoutbursts
and almost lacking normal outbursts
\citep[see e.g.][]{bai79wzsge,dow81wzsge,dow90wxcet},
this definition remained somewhat ambiguous.
Following the dramatic superoutburst of WZ Sge in 2001
\citep{pat02wzsge,ish02wzsgeletter,bab02wzsgeletter},
our understanding of WZ Sge stars has been refreshed.
It was known that periodic variations having
the orbital period were observed in WZ Sge stars
during the early phase of their superoutbursts.
\citet{pat81wzsge} considered that they are orbital humps
and and that they reflect a greatly enhanced mass-transfer
from the secondary.  At that time, the only example was
the 1978--1979 superoutburst of WZ Sge \citep{pat81wzsge}.
The number of objects showing this feature gradually
increased: HV Vir in 1992 \citep{kat01hvvir} [although
\citet{bar92hvvir,men92hvviriauc,lei94hvvir} reported
the same feature, they considered the variations
to be usual superhumps] and AL Com in 1995
\citep{kat96alcom,pat96alcom,how96alcom,nog97alcom}.
Following the outburst of AL Com, the presence of
double-wave modulations during the early stage of
the superoutburst was established.  These variations
are currently referred to as early superhumps.
\citet{osa02wzsgehump} identified early superhumps as
the manifestation of the 2:1 resonance, in addition to
the 3:1 resonance which causes superhumps and
superoutbursts in SU UMa stars
\citep{whi88tidal,osa89suuma,hir90SHexcess,lub91SHa}
and in WZ Sge stars later during superoutbursts.
WZ Sge stars are currently defined as dwarf novae in which
the 2:1 resonance plays a role during their superoutbursts
\citep{kat15wzsge}.

\begin{figure*}
  \begin{center}
    \includegraphics[height=8cm]{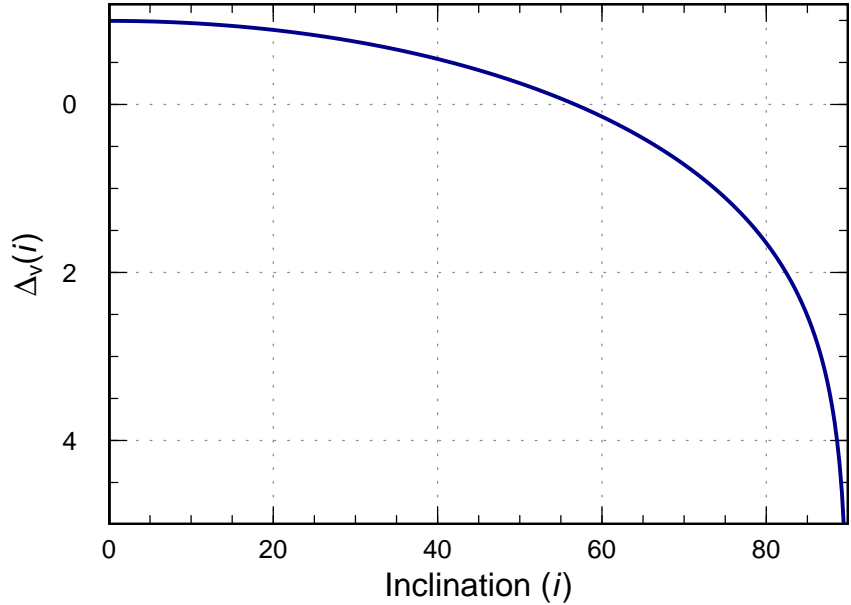}
  \end{center}
  \caption{Dependence of the apparent brightness of
    an optically thick disk on the orbital inclination
    using equation (\ref{equ:incldepend}).
    Fainter magnitudes are displayed lower.}
  \label{fig:icorr}
\end{figure*}

   In \citet{kat15wzsge}, I suggested in its subsection
7.10 that the brightness of WZ Sge stars when ordinary
superhumps appear could be used as the standard candle
since the disk is expected to have a size close to
the radius of the 3:1 resonance.
This could be equally applied to SU UMa stars.
There are, however, unavoidable large uncertainties
arising from inclinations.
\citet{pac80ugem} derived a formula for an optically thick
disk (as in dwarf-nova outbursts) considering
the projected area and limb darkening of the disk
\begin{equation}
\label{equ:incldepend}
\Delta M_v(i) = -2.5 \log_{10} \left[ (1 + \frac{3}{2}\cos i) \cos i \right],
\end{equation}
where $\Delta M_v$ and $i$ are the correction to produce
the visual absolute magnitude and the inclination, respectively.
These corrections can become large
\citep{war87CVabsmag,pat11CVdistance}.
This relation is shown in figure \ref{fig:icorr}.
Pole-on systems ($i$=0) are observed 1.0~mag brighter than
the average and systems with $i$=80$^\circ$ are nearly
2~mag fainter.
The use of the accretion disk as the standard candle
is somewhat limited since $i$ is relatively difficult to
measure other than in eclipsing systems.

   In WZ Sge stars, however, the amplitude of early superhumps
is considered to depend on the inclination since early superhumps
arise from a geometric effect (to the observer) of
vertical structures of the disk \citep{osa02wzsgehump}.
\citet{uem12ESHrecon} succeeded
in reproducing the profile of early superhumps by considering
self-eclipse of a vertically extended disk.
\citet{kat15wzsge} used the code by \citet{uem12ESHrecon}
and reasonably reproduced the distribution of the observed
amplitudes of early superhumps except objects with
very large amplitudes (in its subsection 5.5).
The amplitudes of early superhumps could thus be used
instead of $i$.

   At the time of \citet{kat15wzsge}, WZ Sge stars with
known parallaxes were only three (WZ Sge, GW Lib and
V455 And).  Now that Gaia parallaxes are available
\citep[GaiaEDR3:][]{GaiaEDR3}, I calibrated this
``standard candle'' and studied the dependence on the amplitude
of early superhumps.

\section{The Data}

\subsection{Data source}

   The data were mostly taken from table 5 in
\citet{kat15wzsge}.  The quiescent magnitudes were replaced
by average $G$ magnitudes in \citet{GaiaEDR3}.
The amplitudes of early superhumps
were taken from table 2 in \citet{kat15wzsge}.
The full amplitudes of early superhumps (\aesh) correspond
to A2 (mean amplitude) in table 2 in \citet{kat15wzsge}.
Some objects in table 5 in \citet{kat15wzsge} had
more recent superoutbursts.  If the values were improved,
I supplied the data from these superoutbursts.
As stated in \citet{kat15wzsge}, the accuracy of the magnitudes
at which ordinary superhumps appear is $\sim$0.1~mag.
In some cases, zero-point calibrations of unfiltered
CCD magnitudes caused more uncertainties
(particularly in the past).  In such cases,
I used All-Sky Automated Survey
\citep[ASAS-3:][]{ASAS3} $V$ data and
All-Sky Automated Survey for Supernovae (ASAS-SN)
Sky Patrol $V$ data \citep{ASASSN,koc17ASASSNLC}
to obtain well-calibrated magnitudes.
Observations from the AAVSO International Database\footnote{
   $<$http://www.aavso.org/data-download$>$.
} were sometimes used.

   Additional objects were taken from \citet{Pdot9,Pdot10}.
Some objects (mainly since 2017) reported to VSNET \citep{VSNET}
were also included to increase the sample.  The detailed source
of the data of each object is given later in this section.
The objects having 1$\sigma$ errors in Gaia parallaxes
typically less than 20\% were included in the table.
When drawing figures and making statistical analysis,
I limited objects with parallax errors less than 10\%,
corresponding to errors of 0.2~mag.
Dwarf novae whose parallaxes are measured to this accuracy
are nearby objects and I ignored interstellar extinction.
The lack of reddening was confirmed by the blue colors
($BP-RP$) in Gaia magnitudes.
In examining the light curves, I also used Public Data Releases of
the Zwicky Transient Facility \citep{ZTF}
observations\footnote{
   The ZTF data can be obtained from IRSA
$<$https://irsa.ipac.caltech.edu/Missions/ztf.html$>$
using the interface
$<$https://irsa.ipac.caltech.edu/docs/program\_interface/ztf\_api.html$>$
or using a wrapper of the above IRSA API
$<$https://github.com/MickaelRigault/ztfquery$>$.}.

\subsection{Notes on individual objects}

   Simply referring to ``magnitude'', here I mean the magnitude
when ordinary superhumps appeared.

\medskip

\textbf{AL Com}: updated using the 2019 April data \citep{tam21DNespec}.

\textbf{EG Cnc}: updated using the 2018 October data \citep{kim21egcnc}.

\textbf{HV Vir}: the same magnitude was obtained using the 2016 March data
\citep[][and also VSNET data]{ima18hvvirj0120}.

\textbf{RZ Leo}: updated using the 2022 January data
(Outburst detection by Tadashi Kojima.  H. Maehara, vsnet-alert 26522\footnote{
  $<$http://ooruri.kusastro.kyoto-u.ac.jp/mailarchive/vsnet-alert/26522$>$.
};
T. Kato, vsnet-alert 26529\footnote{
  $<$http://ooruri.kusastro.kyoto-u.ac.jp/mailarchive/vsnet-alert/26529$>$.
} and
Y. Tampo, vsnet-alert 26532\footnote{
  $<$http://ooruri.kusastro.kyoto-u.ac.jp/mailarchive/vsnet-alert/26532$>$.
}).  The main observers were Kyoto U. team, Hiroshi Itoh,
Seiichiro Kiyota, Osaka Kyoiku U. team, Shawn Dvorak,
Tam\'as Tordai, Stephen M. Brincat, Vihorlat Observatory team
and Filipp Romanov.

\textbf{UZ Boo}: magnitude updated using snapshot
$V$ measurements.

\textbf{ASAS J102522$-$1542.4}: value updated using ASAS-3 $V$ magnitudes.
The amplitude of early superhumps was taken from \citet{Pdot}.

\textbf{EZ Lyn}: magnitude updated based on $V$ data used in \citet{Pdot}.

\textbf{V455 And}: magnitude updated based on $V$ data used in \citet{Pdot}.

\textbf{OT J111217.4$-$353829}: the epoch of the appearance of ordinary
superhumps was somewhat uncertain.

\textbf{V624 Peg}: the epoch of the appearance of ordinary
superhumps was somewhat uncertain.  The magnitude was
from AAVSO $V$ observations.

\textbf{V1838 Aql}: The amplitude of early superhumps was taken from
\citet{Pdot6}.  See also \citet{ech19v1838aql}.

\textbf{ASASSN-14cv}: early superhumps were reanalyzed in this work.
The magnitude was confirmed during the 2020 July--August superoutburst.

\textbf{V529 Dra}: magnitude updated using snapshot $V$ data.
First case of double superoutburst \citep{Pdot4}
and is one of the best candidates for period bouncers.

\textbf{GS Cet}: data from \citet{Pdot9}.  The magnitude was from
ASAS-SN $V$ data.

\textbf{ASASSN-16eg}: data from \citet{wak17asassn16eg}.
WZ Sge star with an unusually long orbital period.

\textbf{ASASSN-16js}: data from \citet{Pdot9}.  The magnitude was from
ASAS-SN $V$ data.

\textbf{ASASSN-17el}: data from \citet{Pdot10}.
The amplitude of early superhump
was corrected in this work.  The magnitude was from
ASAS-SN $V$ data.

\textbf{PNV J20205397$+$2508145}: data from \citet{Pdot10}.

\textbf{HO Cet}: data from \citet{Pdot}.  The magnitude was from
ASAS-SN $V$ data.

\textbf{V627 Peg}: data from \citet{Pdot2}.

\textbf{PNV J17144255$-$2943481}: data from \Nakataprep ; reanalyzed
in this paper.  This object showed five post-superoutburst
rebrightenings \citep{kat15wzsge}.

\textbf{OV Boo}: data from \citet{ohn19ovboo}.
Population II object \citep{pat17ovboo}.

\textbf{TCP J18154219$+$3515598}: superoutburst in 2017 June\footnote{
  $<$http://www.cbat.eps.harvard.edu/unconf/followups/J18154219$+$3515598.html$>$.
}
(H. Maehara, vsnet-alert 21098\footnote{
  $<$http://ooruri.kusastro.kyoto-u.ac.jp/mailarchive/vsnet-alert/21098$>$.
};
K. Isogai, vsnet-alert 21101\footnote{
  $<$http://ooruri.kusastro.kyoto-u.ac.jp/mailarchive/vsnet-alert/21101$>$.
};
T. Kato, vsnet-alert 21105\footnote{
  $<$http://ooruri.kusastro.kyoto-u.ac.jp/mailarchive/vsnet-alert/21105$>$.
};
spectrum by P. Berardi\footnote{
  $<$http://quasar.teoth.it/html/spectra/tcpj18154219$+$3515598\_PB.png$>$.
} and
T. Kato, vsnet-alert 21109\footnote{
  $<$http://ooruri.kusastro.kyoto-u.ac.jp/mailarchive/vsnet-alert/21109$>$.
}).  This object showed 10 post-superoutburst rebrightenings
(June--September), nine of which were announced on VSNET:
R. J. Modic, vsnet-alert 21175\footnote{
  $<$http://ooruri.kusastro.kyoto-u.ac.jp/mailarchive/vsnet-alert/21175$>$.
};
R. J. Modic, vsnet-alert 21224\footnote{
  $<$http://ooruri.kusastro.kyoto-u.ac.jp/mailarchive/vsnet-alert/21224$>$.
};
E. de Miguel, vsnet-alert 21244\footnote{
  $<$http://ooruri.kusastro.kyoto-u.ac.jp/mailarchive/vsnet-alert/21244$>$.
};
H. Maehara, vsnet-alert 21268\footnote{
  $<$http://ooruri.kusastro.kyoto-u.ac.jp/mailarchive/vsnet-alert/21268$>$.
};
T. Kato, vsnet-alert 21278\footnote{
  $<$http://ooruri.kusastro.kyoto-u.ac.jp/mailarchive/vsnet-alert/21278$>$.
};
K. Isogai, vsnet-alert 21302\footnote{
  $<$http://ooruri.kusastro.kyoto-u.ac.jp/mailarchive/vsnet-alert/21302$>$.
};
H. Maehara, vsnet-alert 21322\footnote{
  $<$http://ooruri.kusastro.kyoto-u.ac.jp/mailarchive/vsnet-alert/21322$>$.
};
R. J. Modic, vsnet-alert 21334\footnote{
  $<$http://ooruri.kusastro.kyoto-u.ac.jp/mailarchive/vsnet-alert/21334$>$.
} and
H. Maehara, vsnet-alert 21347\footnote{
  $<$http://ooruri.kusastro.kyoto-u.ac.jp/mailarchive/vsnet-alert/21347$>$.
}.
See also \citet{zub18j1815}.  The amplitude of early superhumps
is based on re-analysis of the data reported to VSNET.
The main observers were Geoff Stone, Tam\'as Tordai,
Enrique de Miguel, Tonny Vanmunster, Lisnyky Observatory team,
Kyoto U. team, Vihorlat Observatory team, Stephen M. Brincat,
Hiroshi Itoh, Rudolf Nov\'ak, Terskol Observatory team,
Seiichiro Kiyota, Kiyoshi Kasai, Roger D. Pickard,
Natalia Katysheva, Ian Miller, Alexandra M. Zubareva,
Javier Ruiz, William Stein, Sergey Yu. Shugarov,
Lewis M. Cook, TSHAO Observatory team and
Crimean Astrophysical Observatory team.

\textbf{ASASSN-17pm}: also known as PNV J05580574$-$0011155\footnote{
  $<$http://www.cbat.eps.harvard.edu/unconf/followups/J05580574$-$0011155.html$>$.
}.  Superoutburst in 2017 November--December
(T. Vanmunster, vsnet-alert 21624\footnote{
  $<$http://ooruri.kusastro.kyoto-u.ac.jp/mailarchive/vsnet-alert/21624$>$.
};
T. Kato, vsnet-alert 21627\footnote{
  $<$http://ooruri.kusastro.kyoto-u.ac.jp/mailarchive/vsnet-alert/21627$>$.
} and
T. Kato, vsnet-alert 21646\footnote{
  $<$http://ooruri.kusastro.kyoto-u.ac.jp/mailarchive/vsnet-alert/21646$>$.
}).  The amplitude of early superhump was re-analyzed in this work.
There was a short rebrightening, a shallow dip and a long
rebrightening.
The main observers were Hiroshi Itoh, Crimean Astrophysical
Observatory team, Franz-Josef Hambsch, Seiichiro Kiyota,
Tonny Vanmunster, Kyoto U. team, Tam\'as Tordai,
Ian Miller and Domenico Licchelli.

\textbf{ASASSN-18do}: the data were not very good around
the appearance of ordinary superhumps.
Superoutburst in 2018 February--March
(T. Vanmunster, vsnet-alert 21906\footnote{
  $<$http://ooruri.kusastro.kyoto-u.ac.jp/mailarchive/vsnet-alert/21906$>$.
};
T. Kato, vsnet-alert 21921\footnote{
  $<$http://ooruri.kusastro.kyoto-u.ac.jp/mailarchive/vsnet-alert/21921$>$.
}).
This object is eclipsing
(T. Kato, vsnet-alert 21933\footnote{
  $<$http://ooruri.kusastro.kyoto-u.ac.jp/mailarchive/vsnet-alert/21933$>$.
}).  The main observers were Tonny Vanmunster,
Hiroshi Itoh, Jochen Pietz, Tam\'as Tordai,
Crimean Astrophysical Observatory team, Sergey Yu. Shugarov
and Kyoto U. team.

\textbf{ASASSN-18wa}: superoutburst in 2018 September--October
(T. Vanmunster, vsnet-alert 22560\footnote{
  $<$http://ooruri.kusastro.kyoto-u.ac.jp/mailarchive/vsnet-alert/22560$>$.
};
Y. Wakamatsu, vsnet-alert 22565\footnote{
  $<$http://ooruri.kusastro.kyoto-u.ac.jp/mailarchive/vsnet-alert/22565$>$.
};
Y. Wakamatsu, vsnet-alert 22591\footnote{
  $<$http://ooruri.kusastro.kyoto-u.ac.jp/mailarchive/vsnet-alert/22591$>$.
}).  There was a short gap in the observations after
the superoutburst and it was unclear whether there was
a post-superoutburst rebrightening.  The ZTF data showed
a smooth fading tail characteristic to a WZ Sge star.
The main observers were Tam\'as Tordai,
Osaka Kyoiku U. team, Tonny Vanmunster, Hiroshi Itoh
and Geoff Stone.

\textbf{ASASSN-19ag}: superoutburst in 2019 January
(T. Vanmunster, vsnet-alert 22927\footnote{
  $<$http://ooruri.kusastro.kyoto-u.ac.jp/mailarchive/vsnet-alert/22927$>$.
}).
Although brightening and growth of ordinary superhumps
were recorded, the profile before the growth of ordinary
superhumps was not similar to that of early superhumps
(T. Kato, vsnet-alert 22937\footnote{
  $<$http://ooruri.kusastro.kyoto-u.ac.jp/mailarchive/vsnet-alert/22937$>$.
}).  I therefore did not give the amplitude of early
superhumps.  The object showed at least one post-superoutburst
rebrightening in the ZTF data.

\textbf{TCP J06373299$-$0935420} = ASASSN-19de: superoutburst in 2019
February--March\footnote{
  $<$http://www.cbat.eps.harvard.edu/unconf/followups/J06373299$-$0935420.html$>$.
}
(spectrum by R.~Leadbeater, vsnet-alert 23014\footnote{
  $<$http://ooruri.kusastro.kyoto-u.ac.jp/mailarchive/vsnet-alert/23014$>$ and \\
  $<$https://britastro.org/specdb/data\_graph.php?obs\_id=3925$>$.
} and
T. Kato, vsnet-alert 23036\footnote{
  $<$http://ooruri.kusastro.kyoto-u.ac.jp/mailarchive/vsnet-alert/23036$>$.
}).  There was one post-superoutburst rebrightening
(T. Kato, vsnet-alert 23083\footnote{
  $<$http://ooruri.kusastro.kyoto-u.ac.jp/mailarchive/vsnet-alert/23083$>$.
}).
The main observers were Osaka Kyoiku U. team, Hiroshi Itoh,
Stephen M. Brincat, Berto Monard, Franz-Josef Hambsch,
Seiichiro Kiyota, Arto Oksanen, Masanori Mizutani,
Tonny Vanmunster and Yasui Sano.  According to vsnet-alert
postings, Gianluca Masi also reported observations but
I could not receive the data and are not included
in this analysis.

\textbf{TCP J05390410$+$4748030} = ASASSN-19hh: superoutburst in 2019
March--April\footnote{
  $<$http://www.cbat.eps.harvard.edu/unconf/followups/J05390410$+$4748030.html$>$.
}
(T. Vanmunster, vsnet-alert 23068\footnote{
  $<$http://ooruri.kusastro.kyoto-u.ac.jp/mailarchive/vsnet-alert/23068$>$.
};
T. Kato, vsnet-alert 23072\footnote{
  $<$http://ooruri.kusastro.kyoto-u.ac.jp/mailarchive/vsnet-alert/23072$>$.
};
T. Kato, vsnet-alert 23120\footnote{
  $<$http://ooruri.kusastro.kyoto-u.ac.jp/mailarchive/vsnet-alert/23120$>$.
} and
T. Kato, vsnet-alert 23124\footnote{
  $<$http://ooruri.kusastro.kyoto-u.ac.jp/mailarchive/vsnet-alert/23124$>$.
}).
The main observers were Tonny Vanmunster, Seiichiro Kiyota,
Hiroshi Itoh, Crimean Astrophysical Observatory team,
Vihorlat Observatory team, Tam\'as Tordai, Stephen M. Brincat,
Ian Miller and Lewis M. Cook.

\textbf{ASASSN-19hl}: superoutburst in 2019 March--April
(D. Denisenko, vsnet-alert 23090\footnote{
  $<$http://ooruri.kusastro.kyoto-u.ac.jp/mailarchive/vsnet-alert/23090$>$.
};
T. Kato, vsnet-alert 23103\footnote{
  $<$http://ooruri.kusastro.kyoto-u.ac.jp/mailarchive/vsnet-alert/23103$>$.
} and
T. Kato, vsnet-alert 23144\footnote{
  $<$http://ooruri.kusastro.kyoto-u.ac.jp/mailarchive/vsnet-alert/23144$>$.
}).  The main observers were Berto Monard, Franz-Josef Hambsch
and Hiroshi Itoh.

\textbf{V3101 Cyg}: data from \citet{tam20v3101cyg}.
The amplitude of early superhumps was re-analyzed in this work.

\textbf{EQ Lyn}: data from \citet{tam21DNespec}.

\textbf{GY Cet}: superoutburst in 2020 July--August
(P. Schmeer, vsnet-alert 24446\footnote{
  $<$http://ooruri.kusastro.kyoto-u.ac.jp/mailarchive/vsnet-alert/24446$>$.
};
T. Kato, vsnet-alert 24486\footnote{
  $<$http://ooruri.kusastro.kyoto-u.ac.jp/mailarchive/vsnet-alert/24486$>$.
} and
T. Kato, vsnet-alert 24498\footnote{
  $<$http://ooruri.kusastro.kyoto-u.ac.jp/mailarchive/vsnet-alert/24498$>$.
}).  No post-superoutburst rebrightening was present.
The main observers were Franz-Josef Hambsch and Berto Monard.

\textbf{VX For}: superoutburst in 2021 January--February
(Y. Maeda, vsnet-alert 25264\footnote{
  $<$http://ooruri.kusastro.kyoto-u.ac.jp/mailarchive/vsnet-alert/25264$>$.
};
P. Schmeer, vsnet-alert 25265\footnote{
  $<$http://ooruri.kusastro.kyoto-u.ac.jp/mailarchive/vsnet-alert/25265$>$.
} and
Y. Tampo, vsnet-alert 25288\footnote{
  $<$http://ooruri.kusastro.kyoto-u.ac.jp/mailarchive/vsnet-alert/25288$>$.
}).  There were five post-superoutburst rebrightenings
(P. Schmeer, vsnet-alert 25358\footnote{
  $<$http://ooruri.kusastro.kyoto-u.ac.jp/mailarchive/vsnet-alert/25358$>$.
};
P. Schmeer, vsnet-alert 25386\footnote{
  $<$http://ooruri.kusastro.kyoto-u.ac.jp/mailarchive/vsnet-alert/25386$>$.
};
R. Stubbings, vsnet-outburst 26651\footnote{
  $<$http://ooruri.kusastro.kyoto-u.ac.jp/mailarchive/vsnet-outburst/26651$>$.
};
P. Schmeer, vsnet-alert 25414\footnote{
  $<$http://ooruri.kusastro.kyoto-u.ac.jp/mailarchive/vsnet-alert/25414$>$.
} and ASAS-SN $g$=14.3 on 2021 February 28).
This object also showed five post-superoutburst rebrightenings
in 2009 \citep{Pdot2} and has been suggested as
a period bouncer \citep{kat22stageA}.
The main observers were Hiroshi Itoh, Berto Monard, Peter Nelson
and Franz-Josef Hambsch.

\textbf{ASASSN-21et} = TCP J06154200$-$2756220:
superoutburst in 2021 April-May\footnote{
  $<$http://www.cbat.eps.harvard.edu/unconf/followups/J06154200$-$2756220.html$>$.
}
(P. Schmeer, vsnet-alert 25637\footnote{
  $<$http://ooruri.kusastro.kyoto-u.ac.jp/mailarchive/vsnet-alert/25637$>$.
};
T. Kato, vsnet-alert 25658\footnote{
  $<$http://ooruri.kusastro.kyoto-u.ac.jp/mailarchive/vsnet-alert/25658$>$.
} and
T. Kato, vsnet-alert 25780\footnote{
  $<$http://ooruri.kusastro.kyoto-u.ac.jp/mailarchive/vsnet-alert/25780$>$.
}).  There was no post-superoutburst rebrightening in
the ASAS-SN data.
The main observers were Berto Monard and Franz-Josef Hambsch.

   I only deal with statistics in this paper and
the full analysis and figures for these unpublished
objects are planned to appear in separate paper(s).

\section{Result and Discussion}

\subsection{Absolute magnitudes when ordinary superhumps appear}

   The absolute magnitudes when ordinary superhumps appear
[hereafter \mvsh]
are listed in table \ref{tab:magtab}.  Note that this table
includes objects with relative large errors in parallax.
For example, the bright \mvsh\, in HO Cet is likely
a result of the uncertainty in the parallax.
In this table, I gave 1$\sigma$ errors estimated from
the errors in parallax.  They are the main cause of
the overall errors of \mvsh.

\begin{center}
\begin{longtable}{lcccccccc}
\caption{Brightness when superhumps appear}\label{tab:magtab} \\
\hline\hline
Object & Year & $P$\commenta & Mag1\commentb & Mag2\commentc & \aesh &$\varpi$\commentd & $\varpi_{\rm error}$ & \mvsh\commente \\
\hline
\endfirsthead
\caption{Brightness when superhumps appear (continued).} \\
\hline\hline
Object & Year & $P$\commenta & Mag1\commentb & Mag2\commentc & \aesh &$\varpi$\commentd & $\varpi_{\rm error}$ & \mvsh\commente \\
\hline
\endhead
\hline
  \multicolumn{8}{l}{\commenta Orbital or superhump (with a suffix `s') period (d).} \\
  \multicolumn{8}{l}{\commentb Quiescent brightness.} \\
  \multicolumn{8}{l}{\commentc Brightness when ordinary superhumps appear.} \\
  \multicolumn{8}{l}{\commentd Gaia EDR3 parallax (mas).} \\
  \multicolumn{8}{l}{\commente Absolute magnitude when ordinary superhumps appear.  The 1$\sigma$ error estimated from $\varpi_{\rm error}$ is given.} \\
  \multicolumn{8}{l}{} \\
\endfoot
  \multicolumn{8}{l}{\commenta Orbital or superhump (with a suffix `s') period (d).} \\
  \multicolumn{8}{l}{\commentb Quiescent brightness.} \\
  \multicolumn{8}{l}{\commentc Brightness when ordinary superhumps appear.} \\
  \multicolumn{8}{l}{\commentd Gaia EDR3 parallax (mas).} \\
  \multicolumn{8}{l}{\commente Absolute magnitude when ordinary superhumps appear.  The 1$\sigma$ error estimated from $\varpi_{\rm error}$ is given.} \\
  \multicolumn{8}{l}{} \\
\endlastfoot
\xxinput{magtab.inc}
\hline
\end{longtable}
\end{center}

\begin{figure*}
  \begin{center}
    \includegraphics[height=8cm]{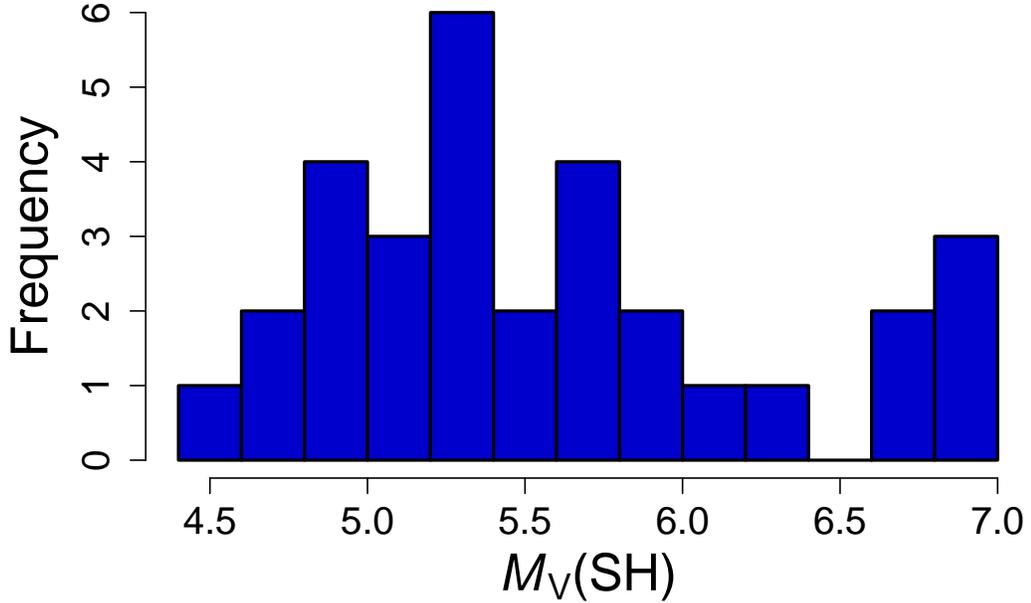}
  \end{center}
  \caption{Distribution of absolute magnitudes when ordinary superhumps
     appear [\mvsh].  Objects with parallax errors
     less than 10\% were selected.}
  \label{fig:mvhist}
\end{figure*}

   The distribution of absolute magnitudes when ordinary superhumps
appear is shown in figure \ref{fig:mvhist}.
The mean vaule is $\langle$\mvsh$\rangle$=$+$5.57 and
the standard deviation is 0.67~mag.  This standard deviation
is too large to be directly used as the standard candle.
Furthermore, the mean vaule is not adequate since \mvsh\,
does not follow a Gaussian distribution as explained later.
The dispersion in \mvsh\, is largely due to the inclination
effect as seen in equation (\ref{equ:incldepend}).

\subsection{Correlation with the amplitude of early superhumps}

\begin{figure*}
  \begin{center}
    \includegraphics[height=11cm]{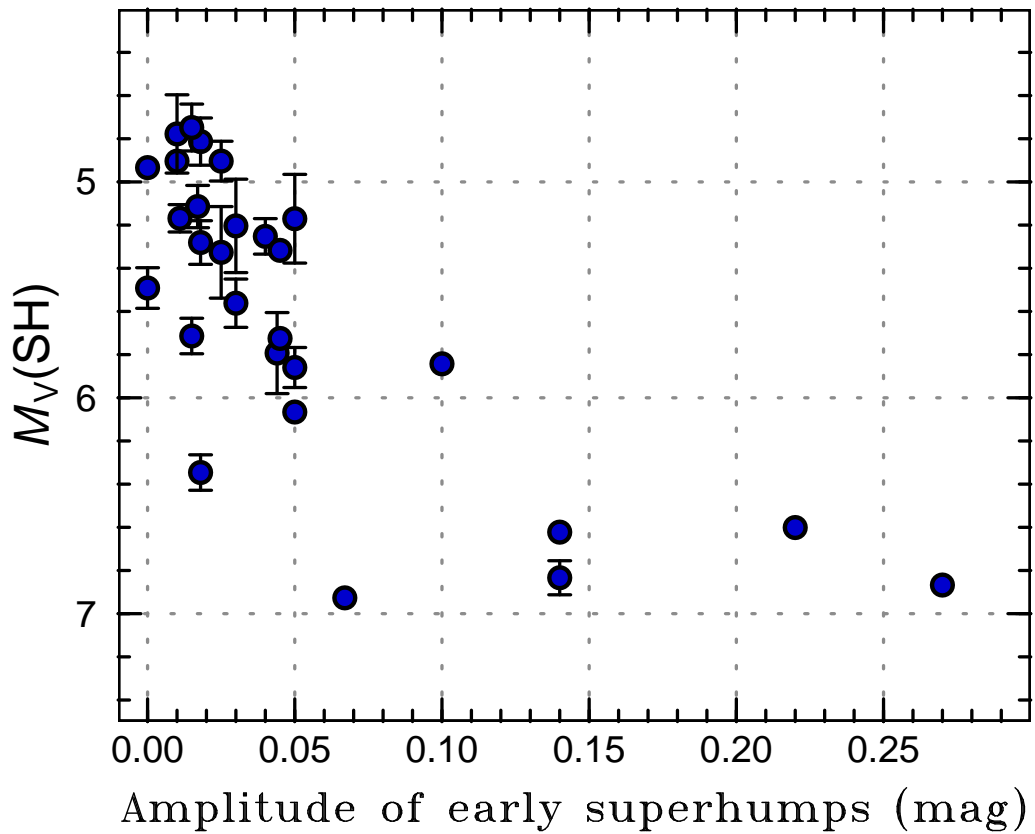}
  \end{center}
  \caption{Dependence of absolute magnitudes when ordinary superhumps
     appear [\mvsh] on amplitudes of early superhumps.
     Objects with parallax errors less than 10\% were selected.
     The error bars reflect the errors from parallax measurements.}
  \label{fig:mvaesh}
\end{figure*}

\begin{figure*}
  \begin{center}
    \includegraphics[height=11cm]{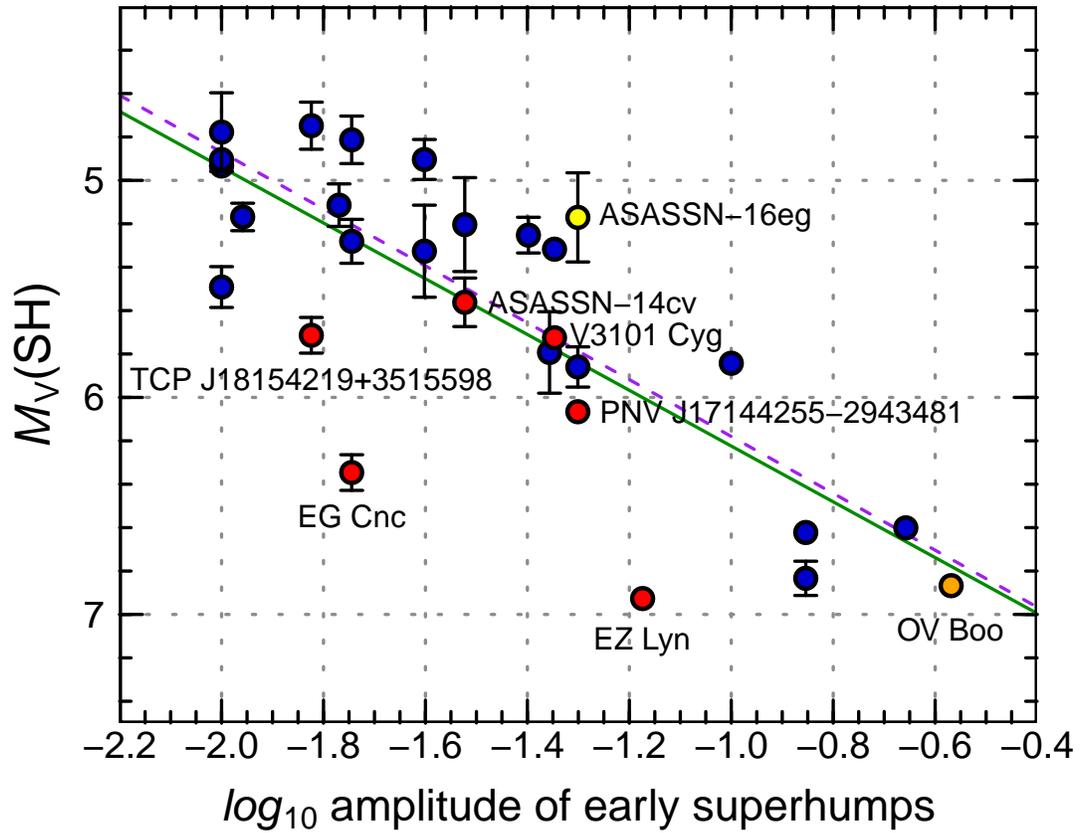}
  \end{center}
  \caption{Dependence of absolute magnitudes when ordinary superhumps
     appear [\mvsh] on amplitudes of early superhumps
     in logarithmic scale.
     Objects with parallax errors less than 10\% were selected.
     Several unusual objects are plotted with different colors.
     Red marks represent objects with multiple post-superoutburst
     rebrightenings.
     (See text for the details).
     The solid and dashed lines represent
     equations (\ref{equ:eshmv}) and (\ref{equ:eshmvrev}),
     respectively.
     The error bars reflect the errors from parallax measurements.}
  \label{fig:mvaeshlog}
\end{figure*}

   Considering that \aesh\, is
expected to be a function of $i$, the relation between
\aesh\, and \mvsh\, is shown
in figure \ref{fig:mvaesh}.  Systems with large \aesh\,
have fainter \mvsh\, as expected.
The relation becomes linear using $\log$ \aesh\,
(figure \ref{fig:mvaeshlog}).  The solid line in the figure
corresponds to
\begin{equation}
M_V(\rm SH) = 7.5(3) + 1.28(20)\log_{10} A_{\rm ESH}.
\label{equ:eshmv}
\end{equation}
In drawing the figure and making a regression,
\mvsh =0.01 was given for objects with \mvsh\ $<$0.01
(The value of 0.01~mag is a realistic upper limit
of the detection of early superhumps; use the same value
in applying the formula to WZ Sge stars without detectable
early superhumps).
The standard deviation from this relation is 0.42~mag.

\subsection{Special objects and the updated relation}

   The two most deviating objects were EG Cnc
(1.1~mag fainter than this relation) and
EZ Lyn (0.9~mag fainter than this relation).
One of the possible reason for EG Cnc is that
the observations of EG Cnc did not start early enough
\citep{kim21egcnc} and \aesh\, might have been underestimated.
EZ Lyn is an eclipsing system and is suggested to be
a period bouncer \citep{zha08j0804,pav07j0804,kat09j0804,
kat13qfromstageA,ama21ezlyn}.

   Although TCP J18154219$+$3515598 with 10 rebrightenings
is also 0.5~mag fainter than this relation,
ASASSN-14cv with 5--8 rebrightenings
[8 in 2014 \citep{skl16asassn14cv,kat15wzsge};
5 in 2020, VSNET data] does not show a strong deviation.
QZ Lib with four rebrightenings \citep{Pdot,pal18qzlib}
and PNV J17144255$-$2943481 with five rebrightenings
are also on this relation.
The very unusual object V3101 Cyg, which showed multiple
rebrightenings and superoutbursts following
the initial superoutburst
\citep{tam20v3101cyg,ham21DNrebv3101cyg}, is also
on the relation.
The most extreme period bouncer in this sample
CRTS J122221.6$-$311524
\citep{kat13j1222,neu17j1222,neu18j1222gwlib}
is not in figure \ref{fig:mvaeshlog} since \aesh\,
is unknown, but has an ordinary \mvsh=5.4.

   There is a possible reason of the deviation for
(some) period bouncers if the deviation is real.
They are likely to have small mass ratios
\citep[][and references therein]{kat22stageA}
and the tidal torque on the disk is
expected to be weak.  This may cause a weaker effect
of the 2:1 resonance and \aesh\, may be smaller
than in other WZ Sge stars.  Since a number of WZ Sge stars
with multiple rebrightenings are on the relation,
this explanation does not seem to apply to all
period bouncers.  This possibility needs
to be studied further both from the observational
and theoretical sides.

   The only population II object below the period minimum
OV Boo \citep{lit07j1507,pat08j1507,uth11j1507,pat17ovboo,ohn19ovboo}
is on the relation.  This object is located near
the period minimum of population II cataclysmic variables
are it may not be as unusual as more evolved period bouncers.
ASASSN-16eg is a WZ Sge star with a very long orbital
period \citep{wak17asassn16eg}.  The deviation to the brighter
side may reflect the large disk size, but the error in
the parallax is still large to draw a definite conclusion.

   Disregarding the objects deviating by more than 0.6~mag
from equation (\ref{equ:eshmv}), I obtained
\begin{equation}
M_V(\rm SH) = 7.5(2) + 1.31(15)\log_{10} A_{\rm ESH}.
\label{equ:eshmvrev}
\end{equation}
The standard deviation from this regression is 0.31~mag.
This equation can currently be regarded as the best relation
for WZ Sge stars with potential exceptions of
evolved period bouncers and systems with very long
orbital periods.

\subsection{Standard candle for the average inclination}

   The median value of \mvsh\, for all the objects
is $+$5.38, which can be considered as a value for
the average inclination (=1 radian = 57$^\circ$).
This value can be used as the standard candle if
\aesh\, is unknown, but with a 1$\sigma$ error of
0.67~mag.  Although, errors may be even larger in
high-inclination systems, $i$ can probably be 
estimated using eclipses in such systems and there
would be no need for the use of the median \mvsh.

\subsection{Estimation of the inclination from the amplitude of early superhumps}

\begin{figure*}
  \begin{center}
    \includegraphics[height=8cm]{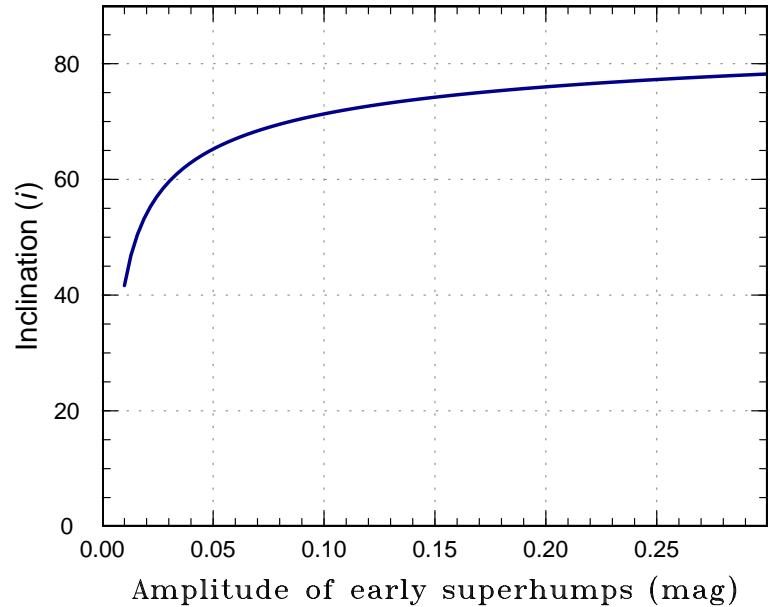}
  \end{center}
  \caption{Relation between the amplitude of early superhumps
    and the inclination.}
  \label{fig:eshtoi}
\end{figure*}

   By equating the equation (\ref{equ:incldepend}) and
the equation (\ref{equ:eshmvrev}) and using the value of
the standard candle of \mvsh=$+$5.38 for $i$=1 radian,
one can obtain an experimental relation between
the amplitude of early superhumps and the inclination
(figure \ref{fig:eshtoi}).  Note that this figure is
based on an assumption that the relation between
$\log$\aesh\, and \mvsh\, is linear.
This figure suggests that \aesh\, is 0.02~mag for
the average inclination (=1 radian).
It is likely that early superhumps are not detectable
for $i <$40$^\circ$.  There is only one object (OV Boo)
with $i >$80$^\circ$ in our sample.  It is possible that
the amplitude of early superhumps is saturated
for very large $i$ and the linear relation would break.
The relation in figure \ref{fig:eshtoi} would be
helpful for estimating $i$ for objects with moderate
\aesh.

\section{Application to MASTER OT J030227.28$+$191754.5}

   MASTER OT J030227.28$+$191754.5 = PNV J03022732$+$1917552
was reported as a possible counterpart \citep{zhi21j0302atel15067}
of IceCube-211125A high-energy neutrino event
\citep{icecube21211125Agcn31126}.
This object had a large outburst amplitude of 10~mag
\citep{zhi21j0302atel15067}.
It was independently discovered by Y. Nakamura\footnote{
  $<$http://www.cbat.eps.harvard.edu/unconf/followups/J03022732$+$1917552.html$>$.
}.
Due to the large outburst amplitude and the possible
association with the neutrino event, it was suspected
to be a nova.  Early spectroscopy indeed showed
narrow emission lines and it was suggested to be
either a narrow-lined He(/N) nova or a dwarf nova
with an exceptionally large amplitude
\citep{tag21j0302atel15072}.  The initial spectrum
did not resemble that of a dwarf nova and
the object was initially favored to be a nova
\citep{pal21j0302atel15073}.  Follow-up observations
did not detect very-high-energy gamma-ray flux
\citep{qui21j0302atel15078,aya21j0302atel15079,
aya21j0302atel15088}.  The optical spectrum on
the second night clarified that the object is
a very high-amplitude dwarf nova
\citep{iso21j0302atel15074}.
CCD images taken 8.5~hr before the neutrino event
indicated that the object was already 1~mag above
the preoutburst level \citep{sar21j0302atel15081}.
Although this observation suggested the association
with the neutrino event less likely, the magnitude
might have been too faint to be considered as
the ignition of the outburst.

   Early superhumps with an amplitude of 0.03~mag
were detected
(T. Kato, vsnet-alert 26477\footnote{
  $<$http://ooruri.kusastro.kyoto-u.ac.jp/mailarchive/vsnet-alert/26477$>$.
}).
The object eventually started to show ordinary
superhumps on 2021 December 30, 30--32~d after
the outburst detection
(T. Kato, vsnet-alert 26501\footnote{
  $<$http://ooruri.kusastro.kyoto-u.ac.jp/mailarchive/vsnet-alert/26501$>$.
}).  This delay of the appearance of ordinary superhumps
was the longest ever observed \citep{kat15wzsge}.

   Due to the unusual nature and the extremely slow
development of ordinary superhumps in this object,
I applied the present method to estimate the distance
and luminosity.
Ordinary superhumps appeared at a magnitude of 14.9
(VSNET data; \Tampoprep).
Using the equation (\ref{equ:eshmvrev}),
\mvsh\, is expected to be $+$5.6 and the distance modulus
is estimated to be 9.3, which corresponds to $\sim$720~pc.
The SDSS magnitude $g$=22.0 \citep{SDSS7} corresponds to
the quiescent absolute magnitude of $+$12.7, which is
one of the faintest recorded in WZ Sge stars
\citep{tam20v3101cyg}.
The peak $V$ magnitude 11.8 corresponds to $M_V$=$+$2.6,
which is also among the brightest \citep{tam20v3101cyg}.

   To examine the possibility if any nuclear reaction was
involved in the outburst of this object, I made a comparison with
the faintest measurement of an outbursting nova.
The recurrent nova T Pyx was recorded on the rise at
a magnitude of 13.0 during the 2011 eruption
by M. Linnolt \citep{waa11tpyxiauc9205,sch13tpyx}
[the initial spectrum was taken several hours later
by \citet{ara15tpyx}, confirming the nova eruption].
This corresponds to $M_V$=$+$0.7 and
the maximum $M_V$ of MASTER OT J030227.28$+$191754.5
is still 6 times below this value.
The faint brightness of T Pyx during the rise was probably
due to the large bolometric correction and the actual
upper limit of the luminosity of the white dwarf
due to nuclear reaction should be 
much lower than this estimate (compared to T Pyx)
based on the $V$-band data.

\section*{Acknowledgements}

This work was supported by JSPS KAKENHI Grant Number 21K03616.
This research has made use of the AAVSO Variable Star Index
(VSX, \cite{wat06VSX}) and NASA's Astrophysics Data System.
I am deeply indebted to world-wide observers who reported
observations, outburst detection and information to
VSNET, VSOLJ and AAVSO.
The author is grateful to the ASAS-3 and
ASAS-SN teams for making their data available to the public.
We acknowledge with thanks the variable star
observations from the AAVSO International Database contributed
by observers worldwide and used in this research.
The author is grateful to Naoto Kojiguchi for helping
downloading the ZTF data.

Based on observations obtained with the Samuel Oschin 48-inch
Telescope at the Palomar Observatory as part of
the Zwicky Transient Facility project. ZTF is supported by
the National Science Foundation under Grant No. AST-1440341
and a collaboration including Caltech, IPAC, 
the Weizmann Institute for Science, the Oskar Klein Center
at Stockholm University, the University of Maryland,
the University of Washington, Deutsches Elektronen-Synchrotron
and Humboldt University, Los Alamos National Laboratories, 
the TANGO Consortium of Taiwan, the University of 
Wisconsin at Milwaukee, and Lawrence Berkeley National Laboratories.
Operations are conducted by COO, IPAC, and UW.

The ztfquery code was funded by the European Research Council
(ERC) under the European Union's Horizon 2020 research and 
innovation programme (grant agreement n$^{\circ}$759194
-- USNAC, PI: Rigault).

\section*{List of objects in this paper}

\xxinput{objlist.inc}

\section*{References}

  I provide two forms of the references section (for ADS
and as published) so that the references can be easily
incorporated into ADS.

\renewcommand\refname{\textbf{References (for ADS)}}

\newcommand{\noop}[1]{}\newcommand{\hyphalt}{-}

\xxinput{wzmvaph.bbl}

\renewcommand\refname{\textbf{References (as published)}}

\xxinput{wzmv.bbl.vsolj}


\begin{thebibliography}{}

\bibitem[{Abazajian} et~al.(2009)]{SDSS7}
  {Abazajian}, K.~N., {et~al.}\ 2009, ApJS, 182, 543 (arXiv:0812.0649)

\bibitem[{Amantayeva} et~al.(2021)]{ama21ezlyn}
  {Amantayeva}, A., {Zharikov}, S., {Page}, K.~L., {Pavlenko}, E.,
  {Sosnovskij}, A., {Khokhlov}, S., \& {Ibraimov}, M.\ 2021, ApJ, 918, 58
  (https://doi.org/10.3847/1538-4357/ac0e36)

\bibitem[{Arai} et~al.(2015)]{ara15tpyx}
  {Arai}, A., {Isogai}, M., {Yamanaka}, M., {Akitaya}, H., \& {Uemura}, M.\
  2015, Acta\ Polytechnica\ CTU\ proceedings, 2, 257

\bibitem[{Ayala}(2021a)]{aya21j0302atel15079}
  {Ayala}, H.\ 2021a, Astron.\ Telegram, 15079, 1

\bibitem[{Ayala}(2021b)]{aya21j0302atel15088}
  {Ayala}, H.\ 2021b, Astron.\ Telegram, 15088, 1

\bibitem[{Baba} et~al.(2002)]{bab02wzsgeletter}
  {Baba}, H., {et~al.}\ 2002, PASJ, 54, L7 (arXiv:astro-ph/0112374)

\bibitem[{Bailey}(1979)]{bai79wzsge}
  {Bailey}, J.\ 1979, MNRAS, 189, 41P (https://doi.org/10.1093/mnras/189.1.41P)

\bibitem[{Barwig} et~al.(1992)]{bar92hvvir}
  {Barwig}, H., {Mantel}, K.~H., \& {Ritter}, H.\ 1992, A\&A, 266, L5

\bibitem[{Downes} and {Margon}(1981)]{dow81wzsge}
  {Downes}, R.~A., \& {Margon}, B.\ 1981, MNRAS, 197, 35P
  (https://doi.org/10.1093/mnras/197.1.35P)

\bibitem[{Downes}(1990)]{dow90wxcet}
  {Downes}, R.~A.\ 1990, AJ, 99, 339 (https://doi.org/10.1086/115332)

\bibitem[{Echevarr{\'\i}a} et~al.(2019)]{ech19v1838aql}
  {Echevarr{\'\i}a}, J., {et~al.}\ 2019, Rev.\ Mexicana\ Astron.\ Astrof., 55,
  21 (arXiv:1810.09864)

\bibitem[{Gaia Collaboration} et~al.(2021)]{GaiaEDR3}
  {Gaia Collaboration}, {et~al.}\ 2021, A\&A, 649, A1 (arXiv:2012.01533)

\bibitem[{Hameury} and {Lasota}(2021)]{ham21DNrebv3101cyg}
  {Hameury}, J.-M., \& {Lasota}, J.-P.\ 2021, A\&A, 650, A114
  (arXiv:2104.02952)

\bibitem[{Hirose} and {Osaki}(1990)]{hir90SHexcess}
  {Hirose}, M., \& {Osaki}, Y.\ 1990, PASJ, 42, 135

\bibitem[{Howell} et~al.(1996)]{how96alcom}
  {Howell}, S.~B., {De Young}, J., {Mattei}, J.~A., {Foster}, G., {Szkody}, P.,
  {Cannizzo}, J.~K., {Walker}, G., \& {Fierce}, E.\ 1996, AJ, 111, 2367
  (https://doi.org/10.1086/117970)

\bibitem[{IceCube Collaboration}(2021)]{icecube21211125Agcn31126}
  {IceCube Collaboration}\ 2021, GRB\ Coord.\ Netw.\ Circ., 31126, 1

\bibitem[{Imada} et~al.(2018)]{ima18hvvirj0120}
  {Imada}, A., {Isogai}, K., {Araki}, T., {Tanada}, S., {Yanagisawa}, K., \&
  {Kawai}, N.\ 2018, PASJ, 70, 2 (arXiv:1711.06080)

\bibitem[{Ishioka} et~al.(2002)]{ish02wzsgeletter}
  {Ishioka}, R., {et~al.}\ 2002, A\&A, 381, L41 (arXiv:astro-ph/0111432)

\bibitem[{Isogai} et~al.(2021)]{iso21j0302atel15074}
  {Isogai}, K., {et~al.}\ 2021, Astron.\ Telegram, 15074, 1

\bibitem[{Kato}(2022)]{kat22stageA}
  {Kato}, T.\ 2022, VSOLJ\ Variable\ Star\ Bull., 89, (arXiv:2201.02945)

\bibitem[{Kato}(2015)]{kat15wzsge}
  {Kato}, T.\ 2015, PASJ, 67, 108 (arXiv:1507.07659)

\bibitem[{Kato} et~al.(2014)]{Pdot6}
  {Kato}, T., {et~al.}\ 2014, PASJ, 66, 90 (arXiv:1406.6428)

\bibitem[{Kato} et~al.(2013a)]{Pdot4}
  {Kato}, T., {et~al.}\ 2013a, PASJ, 65, 23 (arXiv:1210.0678)

\bibitem[{Kato} et~al.(2009a)]{Pdot}
  {Kato}, T., {et~al.}\ 2009a, PASJ, 61, S395 (arXiv:0905.1757)

\bibitem[{Kato} et~al.(2017)]{Pdot9}
  {Kato}, T., {et~al.}\ 2017, PASJ, 69, 75 (arXiv:1706.03870)

\bibitem[{Kato} et~al.(2020)]{Pdot10}
  {Kato}, T., {et~al.}\ 2020, PASJ, 72, 14 (arXiv:1911.04645)

\bibitem[{Kato} et~al.(2010)]{Pdot2}
  {Kato}, T., {et~al.}\ 2010, PASJ, 62, 1525 (arXiv:1009.5444)

\bibitem[{Kato} et~al.(2013b)]{kat13j1222}
  {Kato}, T., {Monard}, B., {Hambsch}, F.-J., {Kiyota}, S., \& {Maehara}, H.\
  2013b, PASJ, 65, L11 (arXiv:1307.5936)

\bibitem[{Kato} et~al.(1996)]{kat96alcom}
  {Kato}, T., {Nogami}, D., {Baba}, H., {Matsumoto}, K., {Arimoto}, J.,
  {Tanabe}, K., \& {Ishikawa}, K.\ 1996, PASJ, 48, L21
  (https://doi.org/10.1093/pasj/48.2.L21)

\bibitem[{Kato} and {Osaki}(2013)]{kat13qfromstageA}
  {Kato}, T., \& {Osaki}, Y.\ 2013, PASJ, 65, 115 (arXiv:1307.5588)

\bibitem[{Kato} et~al.(2009b)]{kat09j0804}
  {Kato}, T., {et~al.}\ 2009b, PASJ, 61, 601 (arXiv:0903.1685)

\bibitem[{Kato} et~al.(2001)]{kat01hvvir}
  {Kato}, T., {Sekine}, Y., \& {Hirata}, R.\ 2001, PASJ, 53, 1191
  (arXiv:astro-ph/0110207)

\bibitem[{Kato} et~al.(2004)]{VSNET}
  {Kato}, T., {Uemura}, M., {Ishioka}, R., {Nogami}, D., {Kunjaya}, C., {Baba},
  H., \& {Yamaoka}, H.\ 2004, PASJ, 56, S1 (arXiv:astro-ph/0310209)

\bibitem[{Kimura} et~al.(2021)]{kim21egcnc}
  {Kimura}, M., {et~al.}\ 2021, PASJ, 73, 1 (arXiv:2008.11871)

\bibitem[{Kochanek} et~al.(2017)]{koc17ASASSNLC}
  {Kochanek}, C.~S., {et~al.}\ 2017, PASP, 129, 104502 (arXiv:1706.07060)

\bibitem[{Leibowitz} et~al.(1994)]{lei94hvvir}
  {Leibowitz}, E.~M., {Mendelson}, H., {Bruch}, A., {Duerbeck}, H.~W.,
  {Seitter}, W.~C., \& {Richter}, G.~A.\ 1994, ApJ, 421, 771
  (https://doi.org/10.1086/173689)

\bibitem[{Littlefair} et~al.(2007)]{lit07j1507}
  {Littlefair}, S.~P., {Dhillon}, V.~S., {Marsh}, T.~R., {G{\"a}nsicke}, B.~T.,
  {Baraffe}, I., \& {Watson}, C.~A.\ 2007, MNRAS, 381, 827 (arXiv:0708.0097)

\bibitem[{Lubow}(1991)]{lub91SHa}
  {Lubow}, S.~H.\ 1991, ApJ, 381, 259 (https://doi.org/10.1086/170647)

\bibitem[{Masci} et~al.(2019)]{ZTF}
  {Masci}, F.-J., {et~al.}\ 2019, PASP, 131, 018003 (arXiv:1902.01872)

\bibitem[{Mendelson} et~al.(1992)]{men92hvviriauc}
  {Mendelson}, H., {Leibowitz}, E.~M., {Brosch}, N., \& {Almoznino}, E.\ 1992,
  IAU\ Circ., 5509

\bibitem[{Neustroev} et~al.(2017)]{neu17j1222}
  {Neustroev}, V.~V., {et~al.}\ 2017, MNRAS, 467, 597 (arXiv:1701.03134)

\bibitem[{Neustroev} et~al.(2018)]{neu18j1222gwlib}
  {Neustroev}, V.~V., {et~al.}\ 2018, A\&A, 611, A13 (arXiv:1712.03515)

\bibitem[{Nogami} et~al.(1997)]{nog97alcom}
  {Nogami}, D., {Kato}, T., {Baba}, H., {Matsumoto}, K., {Arimoto}, J.,
  {Tanabe}, K., \& {Ishikawa}, K.\ 1997, ApJ, 490, 840
  (https://doi.org/10.1086/304881)

\bibitem[{Ohnishi} et~al.(2019)]{ohn19ovboo}
  {Ohnishi}, R., {et~al.}\ 2019, PASJ,  submitted

\bibitem[{Osaki}(1989)]{osa89suuma}
  {Osaki}, Y.\ 1989, PASJ, 41, 1005

\bibitem[{Osaki} and {Meyer}(2002)]{osa02wzsgehump}
  {Osaki}, Y., \& {Meyer}, F.\ 2002, A\&A, 383, 574 (arXiv:astro-ph/0112309)

\bibitem[{Paczy\'{n}ski} and {Schwarzenberg-Czerny}(1980)]{pac80ugem}
  {Paczy\'{n}ski}, B., \& {Schwarzenberg-Czerny}, A.\ 1980, Acta\ Astron., 30,
  127

\bibitem[{Pala} et~al.(2018)]{pal18qzlib}
  {Pala}, A.~F., {Schmidtobreick}, L., {Tappert}, C., {G{\"a}nsicke}, B.~T., \&
  {Mehner}, A.\ 2018, MNRAS, 481, 2523 (arXiv:1809.02135)

\bibitem[{Paliya}(2021)]{pal21j0302atel15073}
  {Paliya}, V.~S.\ 2021, Astron.\ Telegram, 15073, 1

\bibitem[{Patterson} et~al.(1981)]{pat81wzsge}
  {Patterson}, J., {McGraw}, J.~T., {Coleman}, L., \& {Africano}, J.~L.\ 1981,
  ApJ, 248, 1067 (https://doi.org/10.1086/159236)

\bibitem[{Patterson}(2011)]{pat11CVdistance}
  {Patterson}, J.\ 2011, MNRAS, 411, 2695 (arXiv:0903.1006)

\bibitem[{Patterson} et~al.(1996)]{pat96alcom}
  {Patterson}, J., {Augusteijn}, T., {Harvey}, D.~A., {Skillman}, D.~R.,
  {Abbott}, T. M.~C., \& {Thorstensen}, J.\ 1996, PASP, 108, 748
  (https://doi.org/10.1086/133798)

\bibitem[{Patterson} et~al.(2017)]{pat17ovboo}
  {Patterson}, J., {et~al.}\ 2017, Society\ for\ Astronom.\ Sciences\ Ann.\
  Symp., 36, 1

\bibitem[{Patterson} et~al.(2002)]{pat02wzsge}
  {Patterson}, J., {et~al.}\ 2002, PASP, 114, 721 (arXiv:astro-ph/0204126)

\bibitem[{Patterson} et~al.(2008)]{pat08j1507}
  {Patterson}, J., {Thorstensen}, J.~R., \& {Knigge}, C.\ 2008, PASP, 120, 510
  (arXiv:0803.3548)

\bibitem[{Pavlenko} et~al.(2007)]{pav07j0804}
  {Pavlenko}, E., {et~al.}\ 2007, in ASP\ Conf.\ Ser.\ 372, 15th European
  Workshop on White Dwarfs, ed. R. {Napiwotzki}, \& M.~R. {Burleigh}  (San
  Francisco: ASP) p.~511 (arXiv:0712.1956)

\bibitem[{Pojma\'nski}(2002)]{ASAS3}
  {Pojma\'nski}, G.\ 2002, Acta\ Astron., 52, 397 (arXiv:astro-ph/0210283)

\bibitem[{Quinn} et~al.(2021)]{qui21j0302atel15078}
  {Quinn}, J., {VERISTAS Collaboration}, {Metzger}, B., \& {Sokoloski}, J.\
  2021, Astron.\ Telegram, 15078, 1

\bibitem[{Sarneczky} et~al.(2021)]{sar21j0302atel15081}
  {Sarneczky}, K., {Vinko}, J., \& {Kiss}, L.\ 2021, Astron.\ Telegram, 15081,
  1

\bibitem[{Schaefer} et~al.(2013)]{sch13tpyx}
  {Schaefer}, B.~E., {et~al.}\ 2013, ApJ, 773, 55 (arXiv:1109.0065)

\bibitem[{Shappee} et~al.(2014)]{ASASSN}
  {Shappee}, B.~J., {et~al.}\ 2014, ApJ, 788, 48 (arXiv:1310.2241)

\bibitem[{Sklyanov} et~al.(2016)]{skl16asassn14cv}
  {Sklyanov}, A.~S., {Pavlenko}, E.~P., {Antonyuk}, O.~I., {Antonyuk}, K.~A.,
  {Sosnovsky}, A.~A., {Galeev}, A.~I., {Pit'}, N.~V., \& {Babina}, Y.~V.\ 2016,
  Astrophys.\ Bull., 71, 293 (https://doi.org/10.1134/S1990341316030044)

\bibitem[{Taguchi} et~al.(2021)]{tag21j0302atel15072}
  {Taguchi}, K., {Shibata}, M., {Masayuki}, Y., {Isogai}, K., {Tampo}, Y.,
  {Kojiguchi}, N., {Ito}, J., \& {Kato}, T.\ 2021, Astron.\ Telegram, 15072, 1

\bibitem[{Tampo} et~al.(2021)]{tam21DNespec}
  {Tampo}, Y., {et~al.}\ 2021, PASJ, 73, 753 (arXiv:2104.04948)

\bibitem[{Tampo} et~al.(2020)]{tam20v3101cyg}
  {Tampo}, Y., {et~al.}\ 2020, PASJ, 72, 49 (arXiv:2004.10508)

\bibitem[{Uemura} et~al.(2012)]{uem12ESHrecon}
  {Uemura}, M., {Kato}, T., {Ohshima}, T., \& {Maehara}, H.\ 2012, PASJ, 64, 92
  (arXiv:1203.1358)

\bibitem[{Uthas} et~al.(2011)]{uth11j1507}
  {Uthas}, H., {Knigge}, C., {Long}, K.~S., {Patterson}, J., \& {Thorstensen},
  J.\ 2011, MNRAS, 414, L85 (arXiv:1104.1180)

\bibitem[{Waagan} et~al.(2011)]{waa11tpyxiauc9205}
  {Waagan}, E., {Linnolt}, M., \& {Pearce}, A.\ 2011, IAU\ Circ., 9205, 1

\bibitem[{Wakamatsu} et~al.(2017)]{wak17asassn16eg}
  {Wakamatsu}, Y., {et~al.}\ 2017, PASJ, 69, 89 (arXiv:1708.09206)

\bibitem[{Warner}(1995)]{war95book}
  {Warner}, B.\ 1995, Cataclysmic Variable Stars (Cambridge: Cambridge
  University Press)

\bibitem[{Warner}(1987)]{war87CVabsmag}
  {Warner}, B.\ 1987, MNRAS, 227, 23 (https://doi.org/10.1093/mnras/227.1.23)

\bibitem[{Watson} et~al.(2006)]{wat06VSX}
  {Watson}, C.~L., {Henden}, A.~A., \& {Price}, A.\ 2006, Society\ for\
  Astronom.\ Sciences\ Ann.\ Symp., 25, 47

\bibitem[{Whitehurst}(1988)]{whi88tidal}
  {Whitehurst}, R.\ 1988, MNRAS, 232, 35
  (https://doi.org/10.1093/mnras/232.1.35)

\bibitem[{Zharikov} et~al.(2008)]{zha08j0804}
  {Zharikov}, S.~V., {et~al.}\ 2008, A\&A, 486, 505 (arXiv:0804.1947)

\bibitem[{Zhirkov} et~al.(2021)]{zhi21j0302atel15067}
  {Zhirkov}, K., {et~al.}\ 2021, Astron.\ Telegram, 15067, 1

\bibitem[{Zubareva} et~al.(2018)]{zub18j1815}
  {Zubareva}, A.~M., {Shugarov}, S.~Y., \& {Zharova}, A.~V.\ 2018, in A. A.
  Boyarchuk Memorial Conference, INASAN Science Proceedings, ed. D.~V.
  {Bisikalo}, \& D.~S. {Wiebe}  (Moscow: Online at
  http://www.inasan.ru/wp{\hyphalt}content/uploads/2018/12/Boyarchuk.pdf)
  p.~120 (https://doi.org/10.26087/INASAN.2018.1.1.022)

\end{thebibliography}
\end{document}